\documentclass[10pt]{wlscirep}

\title{Variation of the spin textures of 2-species spin-1 condensates
studied beyond the single spatial mode approximation and the experimental
identification of these textures}

\author[1]{Y. Z. He}
\author[2]{Y. M. Liu}
\author[1,*]{C. G. Bao}
\affil[1]{School of Physics, Sun Yat-Sen University, Guangzhou, 510275, P. R. China}
\affil[2]{Department of physics, Shaoguan University, shaoguan, 510205, P. R. China}
\affil[*]{Corresponding author: C.G. Bao, stsbcg@mail.sysu.edu.cn}

\begin{abstract}
Based on the numerical solutions of the coupled Gross-Pitaevskii equations, the spin-textures of a Bose-Einstein condensate with two kinds of spin-1 atoms have been studied. Besides, the probability densities of an atom in spin component $\mu$ and of two correlated atoms one in $\mu$ and one in $\nu$ have been calculated. By an analysis of the densities, four types of texture have been found. (i) Both species are in polar phase. (ii) Both species are in ferromagnetic (f) phase with all the spins lying along a direction. (iii) Both species are in f phase but the spins of different species are lying along opposite directions. (iv) One species in f phase, one species in quasi-f phase (where the spins are divided into two groups lying along opposite directions). This finding simplifies the previous classification. Moreover, we found that the variation and transition of the spin-textures can be sensitively reflected by these probability densities. Therefore, the theoretical and experimental studies on these densities provide a way to discriminate the spin-textures and/or to determine the parameters (say, the strengths of interaction) involved.
\end{abstract}

\begin{document}

\flushbottom
\maketitle

\section*{Introduction}

It is well known that the Bose-Einstein condensates (BEC) as man-made and controllable systems are promising in application. In particular, due to the optical traps, the spin-degrees of freedom can be liberated and therefore the inherent physics becomes much richer \cite{ref5,ref6,ref7,ref8,ref9,ref10}. The so called spinor BEC (S-BEC) is a very exquisite system. Although the spin-dependent forces are very weak, the spin-textures are decisively determined by them. When the temperature is sufficiently low (say, lower than $10^{-10}$K), the spatial degrees of freedom are nearly frozen and the spin degrees of freedom play essential roles. In this case, the S-BEC might be an ideal system for realizing exquisite control.

The ground state (g.s.) of 1-species S-BEC with spin-1 atoms has two well known phases, namely, the polar and ferromagnetic (f) phases. Due to the influence of the spin-dependent interspecies interactions, the spin-textures of 2-species S-BEC become more complicated. These textures have been studied by a number of authors \cite{ml,ref11,ref12,ref13,ref14,ref15,ref16,ref17,ref18}. In the previous study the single spatial mode approximation has been adopted. The details of the spatial wave functions (wfs) have not yet been taken into account, and the effects of many physical factors are combined together (say, refer to Eq.(1) of \cite{ref14} where the strengths represent a combined effect from the strengths of interaction, the trap frequencies, the compactness of the spatial profile, the number of particles, etc.). Therefore, the results from these studies can not clarify the effect of each parameter input in an experiment. To remedy, the present paper aims also at the 2-species S-BEC but goes beyond the single spatial mode approximation. The coupled Gross-Pitaevskii equations (CGP) have been numerically solved to obtain the exact spatial wfs. Thereby, the results of this paper relate directly to the input parameters, and the effect of each parameter can be clarified.

Furthermore, the probability of an atom lying at a specific spin-component and the probability of two atoms (of the same kind or different kinds) lying at two specific spin-components have been calculated. How these probabilities match the variation of the spin-texture has been studied. Since these probabilities are observables (say, via the time-of-flight images), the spin-texture can be thereby identified via experimental measurements of these probabilities. In particular, emphasis is placed on the study of the critical behavior related to the spin-texture-transitions. It was found that the locations of the critical points depend sensitively on both the spatial wfs and on the spin-dependent force. The former depend essentially on the spin-independent force, which is much stronger than the spin-dependent force. Once the wfs have been known, the observation of the critical behavior can provide useful information on the latter.

In the following derivation, the fractional parentage coefficients for spin-1 many-body systems have been introduced \cite{bao05,bao06}. With these coefficients, all the matrix elements of 1-body and 2-body spin-operators can be straightly obtained without knowing the detailed expressions of the related spin-states. Thus these coefficients developed by the authors facilitate greatly the following derivation.

\section*{The coupled Gross-Pitaevskii equations and the spin-textures}

Let the condensate contain $N_A$ $A$-atoms and $N_B$ $B$-atoms. $N_B/2<N_A<N_B$ is assumed. Let $X$ represents $A$ or $B$. The intra-species interaction for spin-1 atoms is $V_X=\sum_{1\leq i<j\leq N_X}\delta(\mathbf{r}_i-\mathbf{r}_j)(c_{X0}+c_{X2}\hat{\mathbf{F}}_i^X\cdot \hat{\mathbf{F}}_j^X)$, where $\hat{\mathbf{F}}_i^X$ is the spin operator for the $i$-th $X$-atom. When the singlet-pairing force has been ignored, the inter-species interaction is $V_{AB}=\sum_{1\leq i\leq N_A}\sum_{1\leq i'\leq N_B}\delta (\mathbf{r}_i-\mathbf{r}_{i'})(c_{AB0}+c_{AB2}\hat{\mathbf{F}}_i^A\cdot\hat{\mathbf{F}}_{i'}^B)$, In the g.s. all the particles of the same kind will condense to a spatial state (denoted as $\varphi_X$) which is most favorable for binding. Let $\Xi$ be a general normalized total spin-state for both species. Then, the g.s. can be written as
\begin{equation}
 \Psi_{\mathrm{o}}
  =  \prod_{i=1}^{N_A}
     \varphi_A(\mathbf{r}_i)
     \prod_{j=1}^{N_B}
     \varphi_B(\mathbf{r}_j)
     \Xi.
 \label{twf}
\end{equation}

Let $\vartheta_{S_XM_X}^{N_X}$ denotes a normalized and all-symmetric spin-state for the $X$-atoms, where all the spins are coupled to $S_X$ and its $Z$-component $M_X$. According to the theory given in Ref.\citen{katr}, $N_X-S_X$ must be even, $\vartheta_{S_X,M_X}^{N_X}$ is unique when $S_X$ and $M_X$ are given, and the set $\{\vartheta_{S_X,M_X}^{N_X}\}$ (where $S_X=N_X$, $N_X-2$, to 0 or 1) is complete for all-symmetric spin-states with $N_X$ spin-1 atoms. We further introduce $(\vartheta_{S_A}^{N_A}\vartheta_{S_B}^{N_B})_{SM}$ where $S_A$ and $S_B$ are coupled to $S$ and its $Z$-component $M$. The set $\{(\vartheta_{S_A}^{N_A}\vartheta_{S_B}^{N_B})_{SM}\}$ is complete for the total spin-states which is invariant against the permutation of the same kind of spins (but there is no permutation symmetry with respect to the interchange of different kinds of atoms).Since $M$ is the total magnetization of the system, it is fixed and therefore can be omitted from the subscripts. Thus, in general, we have the expansion $\Xi=\sum_{S_AS_BS}D_{S_AS_BS}(\vartheta_{S_A}^{N_A}\vartheta_{S_B}^{N_B})_S$, where $D_{S_AS_BS}$ are the coefficients for expansion.

Let $\frac{1}{2}m_X\omega_X^2r^2$ be an isotropic trap for the $X$-atoms. Let us introduce $m$ and $\omega$, and use $\hbar\omega$ and $\lambda \equiv\sqrt{\hbar/(m\omega)}$ as the units for energy and length throughout this paper. Then, the Hamiltonian is
\begin{equation}
 \hat{H}
  =  \hat{K}_A
    +\hat{K}_B
    +V_A
    +V_B
    +V_{AB},
 \label{h}
\end{equation}
where $\hat{K}_X=\sum_{i=1}^{N_X}\hat{h}_X(i)$, $\hat{h}_X(i)=\frac{1}{2}(-\frac{m}{m_X}\nabla _i^2+\gamma_Xr_i^2)$ and $\gamma_X=\frac{m_X\omega _X^2}{m\omega^2}$.

Let $\hat{\mathbf{S}}_X=\sum_i\hat{\mathbf{F}}_i^X$ be the total spin-operator for the $X$-atoms and $\hat{\mathbf{S}}=\hat{\mathbf{S}}_A+\hat{\mathbf{S}}_B$ be the total spin-operator for both kinds of atoms. Due to the two equalities $\sum_{i<j}\hat{\mathbf{F}}_i^X\cdot \hat{\mathbf{F}}_j^X=\frac{1}{2}(\hat{\mathbf{S}}_X\cdot \hat{\mathbf{S}}_X-2N_X)$ and $\sum_i\sum_j\hat{\mathbf{F}}_i^A\cdot \hat{\mathbf{F}}_j^B=\frac{1}{2}(\hat{\mathbf{S}}\cdot \hat{\mathbf{S}}-\hat{\mathbf{S}}_A\cdot \hat{\mathbf{S}}_A-\hat{\mathbf{S}}_B\cdot\hat{\mathbf{S}}_B)$, and making use of the permutation symmetry inherent in $(\vartheta _{S_A}^{N_A}\vartheta_{S_B}^{N_B})_S$, the Hamiltonian is equivalent to an effective Hamiltonian $H_{\mathrm{eff}}$, in which $V_X$ is replaced by
\begin{equation}
 V_X^{\mathrm{eff}}
  =  \sum_{i<j}
     \delta(\mathbf{r}_i-\mathbf{r}_j)
     [ c_{X0}
      +c_{X2}
       \frac{1}{N_X(N_X-1)}
       (\hat{\mathbf{S}}_X\cdot\hat{\mathbf{S}}_X-2N_X) ],
 \label{vefx}
\end{equation}
and $V_{AB}$ is replaced by
\begin{equation}
 V_{AB}^{\mathrm{eff}}
  =  \sum_{i,j}
     \delta(\mathbf{r}_i-\mathbf{r}_j)
     [ c_{AB0}
      +c_{AB2}\frac{1}{2N_AN_B}
       ( \hat{\mathbf{S}}\cdot\hat{\mathbf{S}}
        -\hat{\mathbf{S}}_A\cdot\hat{\mathbf{S}}_A
        -\hat{\mathbf{S}}_B\cdot \hat{\mathbf{S}}_B ) ].
 \label{vefab}
\end{equation}

With these expressions, we know that $H_{\mathrm{eff}}$ keeps $S_A$, $S_B$, $S$, and $M$ to be conserved. Therefore the g.s. can be rewritten as
\begin{equation}
 \Psi_{\mathrm{o}}
  =  \prod_{i=1}^{N_A}
     \varphi_A(\mathbf{r}_i)
     \prod_{j=1}^{N_B}
     \varphi_B(\mathbf{r}_j)
     (\vartheta_{S_A}^{N_A}\vartheta_{S_B}^{N_B})_S
  \equiv
     \Psi_{S_AS_BS},
 \label{gs}
\end{equation}
where $S_A$, $S_B$, and $S$ together with $\varphi_A$ and $\varphi_B$ have to be determined.

Since the set $(S_AS_BS)$ are good quantum numbers, they can be used to classify the spin-textures (note that $M$ is not necessary in the classification because it manifests only the geometry).

From $\delta(\langle\Psi_{S_AS_BS}|H_{\mathrm{eff}}|\Psi_{S_AS_BS}\rangle/\langle\Psi_{S_AS_BS}|\Psi_{S_AS_BS}\rangle)=0$, we obtain the CGP for $\varphi_A$ and $\varphi_B$ as \cite{ml}
\begin{eqnarray}
 && ( \hat{h}_A
     +\alpha_{11}\varphi_A^2
     +\alpha_{12}\varphi_B^2
     -\varepsilon_A )
    \varphi_A=0,  \label{cgp1} \\
 && ( \hat{h}_B
     +\alpha_{21}\varphi_A^2
     +\alpha_{22}\varphi_B^2
     -\varepsilon_B )
    \varphi_B=0,  \label{cgp2}
\end{eqnarray}
where
\begin{eqnarray*}
 && \alpha_{11}=D_AN_A,\ \ \ 
    \alpha_{12}=D_{AB}N_B,\ \ \ 
    \alpha_{21}=D_{AB}N_A,\ \ \ 
    \alpha_{22}=D_BN_B, \\
 && D_X=c_{X0}+\frac{S_X(S_X+1)-2N_X}{N_X(N_X-1)}c_{X2},\ \ \
    D_{AB}=c_{AB0}+\frac{S(S+1)-S_A(S_A+1)-S_B(S_B+1)}{2N_AN_B}c_{AB2}.
\end{eqnarray*}
$\varphi_A$ and $\varphi_B$ are required to be normalized.

Once the values of $(S_A,S_B,S)$ has been presumed, we can solve the CGP to obtain $\varphi_A$ and $\varphi_B$, together with the total energy
\begin{equation}
 E_{S_AS_BS}
  =  N_A
     \langle
     \varphi_A|
     \hat{h}_A|
     \varphi_A
     \rangle
    +N_B
     \langle
     \varphi_B|
     \hat{h}_B|
     \varphi_B
     \rangle
    +\frac{N_A(N_A-1)}{2}
     D_AI_A
    +\frac{N_B(N_B-1)}{2}
     D_BI_B
    +N_AN_BD_{AB}I_{AB},
 \label{th}
\end{equation}
where $I_X=\int\mathrm{d}\mathbf{r}|\varphi_X(\mathbf{r})|^4$ and $I_{AB}=\int\mathrm{d}\mathbf{r}|\varphi_A(\mathbf{r})\varphi_B(\mathbf{r})|^2$. From a series of presumed $(S_A,S_B,S)$, we can find out the optimal set $(S_A,S_B,S)$ which leads to the minimum of the total energy. This set is the good quantum numbers of the g.s..

\begin{figure}[htbp]
 \centering \resizebox{0.6\columnwidth}{!}{\includegraphics{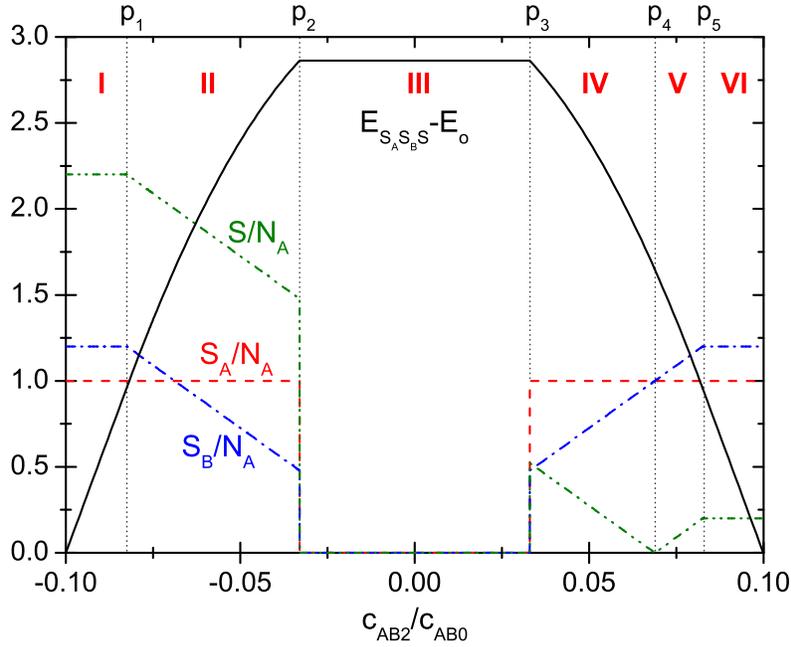} }
 \caption{(color online) Variation of the total energy $E_{S_A S_B S}$ and the spin-texture (specified by $(S_A,S_B,S)$) of the g.s. against $c_{AB2}/c_{AB0}$ from $-0.1$ to $0.1$. The parameters are given as $N_A=1000$, $N_B=1200$, $m_A/m_B=23/87$, $\gamma_A=\gamma_B=1$; $c_{A0}=10^{-3}$, $c_{B0}=2c_{A0}$, $c_{AB0}=0.9c_{A0}$; $c_{A2}=c_{A0}/50$, $c_{B2}=c_{B0}/50$. The units for energy and length are $\hbar\omega$ and $\sqrt{\hbar/(m\omega)}$. The solid line is for $E_{S_AS_BS}-E_{\mathrm{o}}$, where $E_{\mathrm{o}}=3587.52$ is the total energy at the left-end of the curve. At the top of the curve $E_{S_AS_BS}=3590.38$. The ordinates are for $S_A/N_A$, $S_B/N_A$, and $S/N_A$. They are marked by dashed, dash-dot, and dash-dot-dot curves, respectively.}
 \label{fig1}
\end{figure}

\begin{figure}[htbp]
 \centering \resizebox{0.6\columnwidth}{!}{\includegraphics{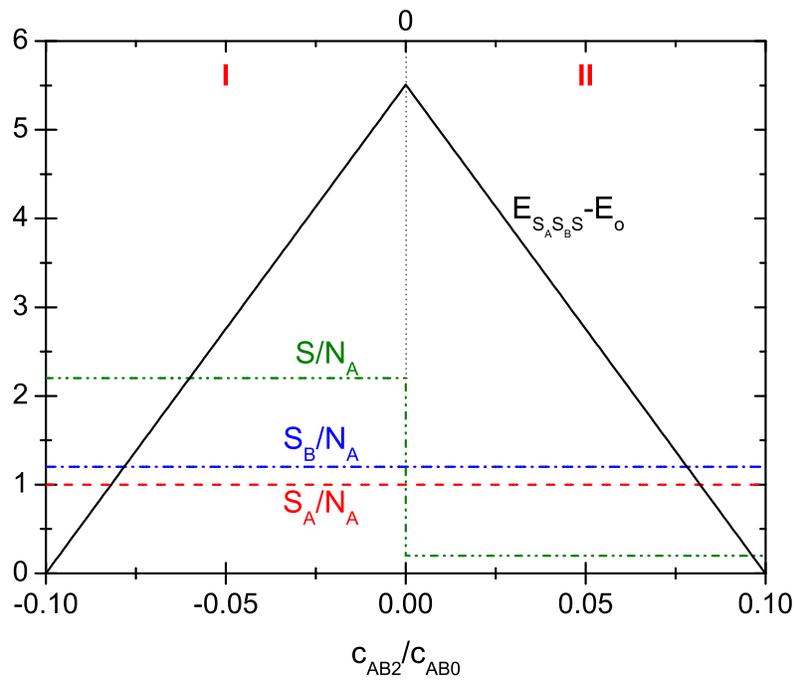}}
 \caption{(color online) The parameters are the same as in Fig.\ref{fig1} but with $c_{A2}=-c_{A0}/50$, $c_{B2}=-c_{B0}/50$. Refer to Fig.1, but with $E_{\mathrm{o}}=3582.25$ and $E_{S_AS_BS}=3587.75$ at the top.}
 \label{fig2}
\end{figure}

\begin{figure}[htbp]
 \centering \resizebox{0.6\columnwidth}{!}{\includegraphics{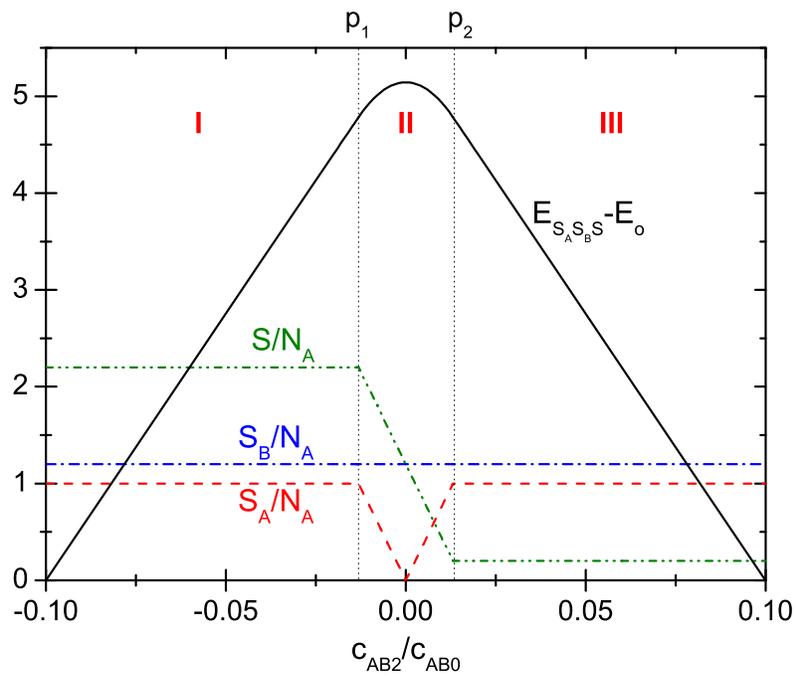}}
 \caption{(color online) The parameters are the same as in Fig.\ref{fig1} but with $c_{B2}=-c_{B0}/50$. Refer to Fig.1, but with $E_{\mathrm{o}}=3582.97$ and $E_{S_AS_BS}=3588.11$ at the top.}
 \label{fig3}
\end{figure}

In this paper, the CGP is solved numerically. Based on the numerical solutions of the CGP, we concentrate on studying the effect of the spin-dependent inter-species interaction on spin-textures. Thus $c_{AB2}$ is considered to be variable. The other parameters are listed in the captions of Fig.\ref{fig1} to Fig.\ref{fig3}. They are so chosen to assure that the g.s. is a miscible state, in which the $A$- and $B$-atoms have a better overlap (otherwise, the effect of $c_{AB2}$ is weak). $N_A\leq N_B$ is assumed. The variation of the spin-texture, specified by $(S_A,S_B,S)$, of the g.s. against $c_{AB2}/c_{AB0}$ is plotted in Fig.\ref{fig1}, \ref{fig2} and \ref{fig3}, respectively, for (i) both $c_{A2}$ and $c_{B2}$ are positive, (ii) both $c_{A2}$ and $c_{B2}$ are negative, and (iii) $c_{A2}>0$ and $c_{B2}<0$. In these figures, the total energy $E_{S_AS_BS}$ is also given.

\begin{figure}[htbp]
 \centering \resizebox{0.6\columnwidth}{!}{\includegraphics{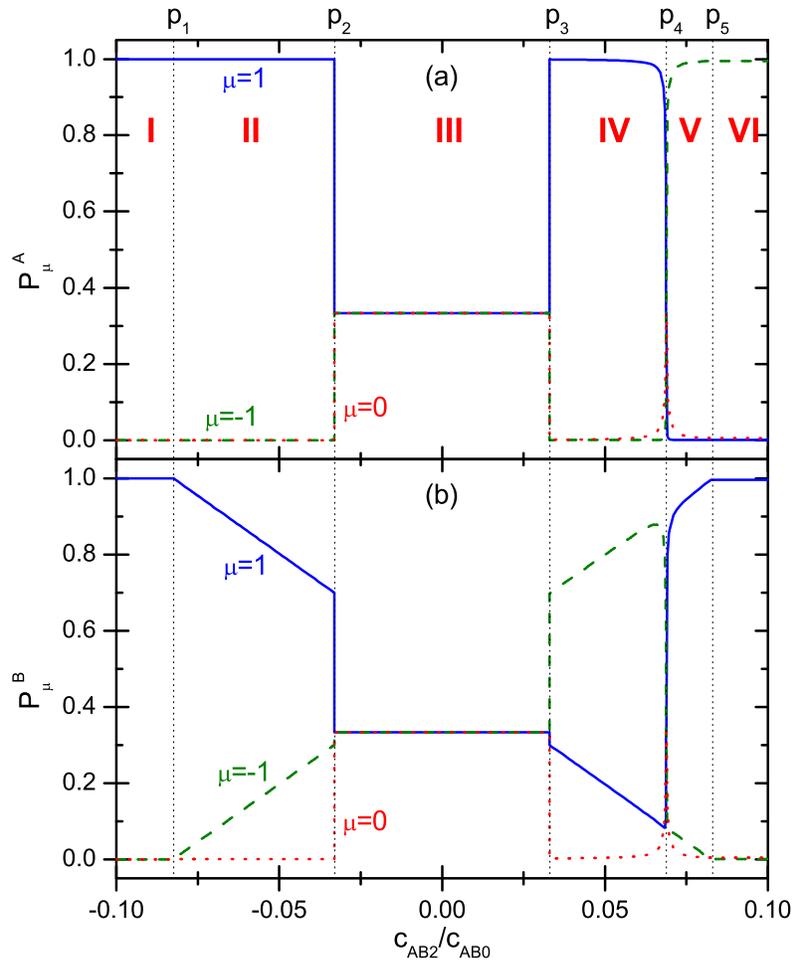}}
 \caption{(color online) $P_{\mu}^A$ (a) and $P_{\mu}^B$ (b) against $c_{AB2}/c_{AB0}$. The parameters are the same as in Fig.\ref{fig1}. Solid line is for $\mu=1$, dotted line is for $\mu=0$, and dash line is for $\mu=-1$. The points $p_1$ to $p_5$ mark the boundary of the zones, refer to Fig.\ref{fig1}.}
 \label{fig4}
\end{figure}

\begin{figure}[htbp]
 \centering \resizebox{0.6\columnwidth}{!}{\includegraphics{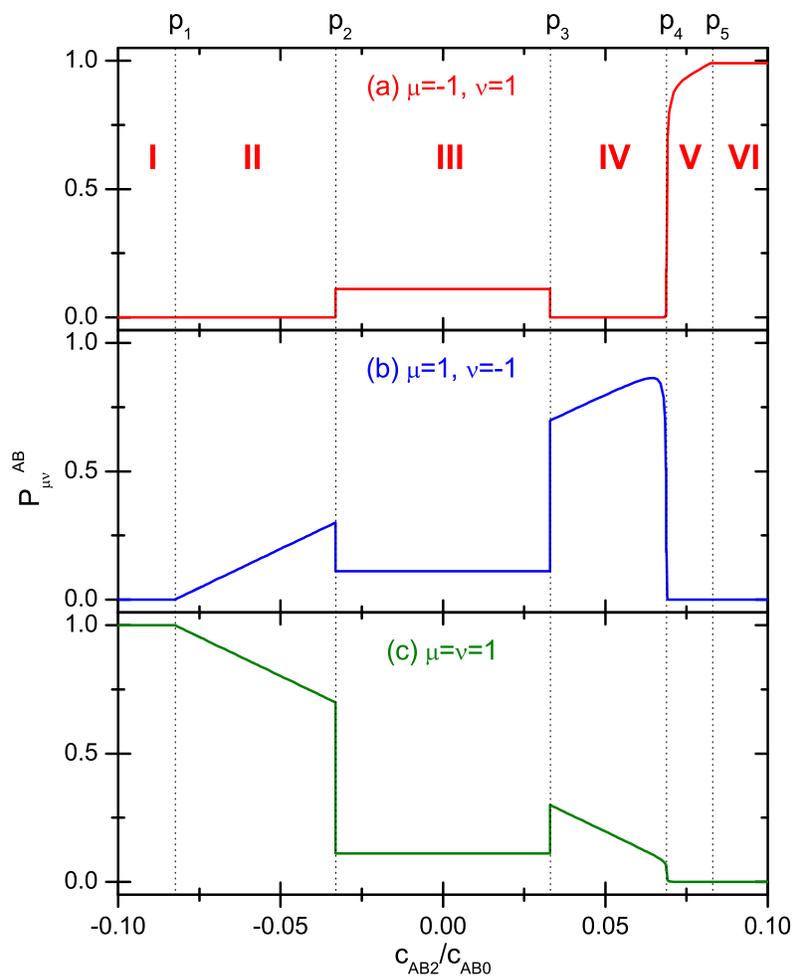}}
 \caption{(color online) Selected $P_{\mu\nu}^{AB}$ against $c_{AB2}/c_{AB0}$. The parameters are the same as in Fig.\ref{fig1}. $\mu$ ($\nu$) is the component of an $A$-atom (a $B$-atom). $\mu$ and $\nu$ are marked in the figure.}
 \label{fig5}
\end{figure}

\begin{figure}[htbp]
 \centering \resizebox{0.6\columnwidth}{!}{\includegraphics{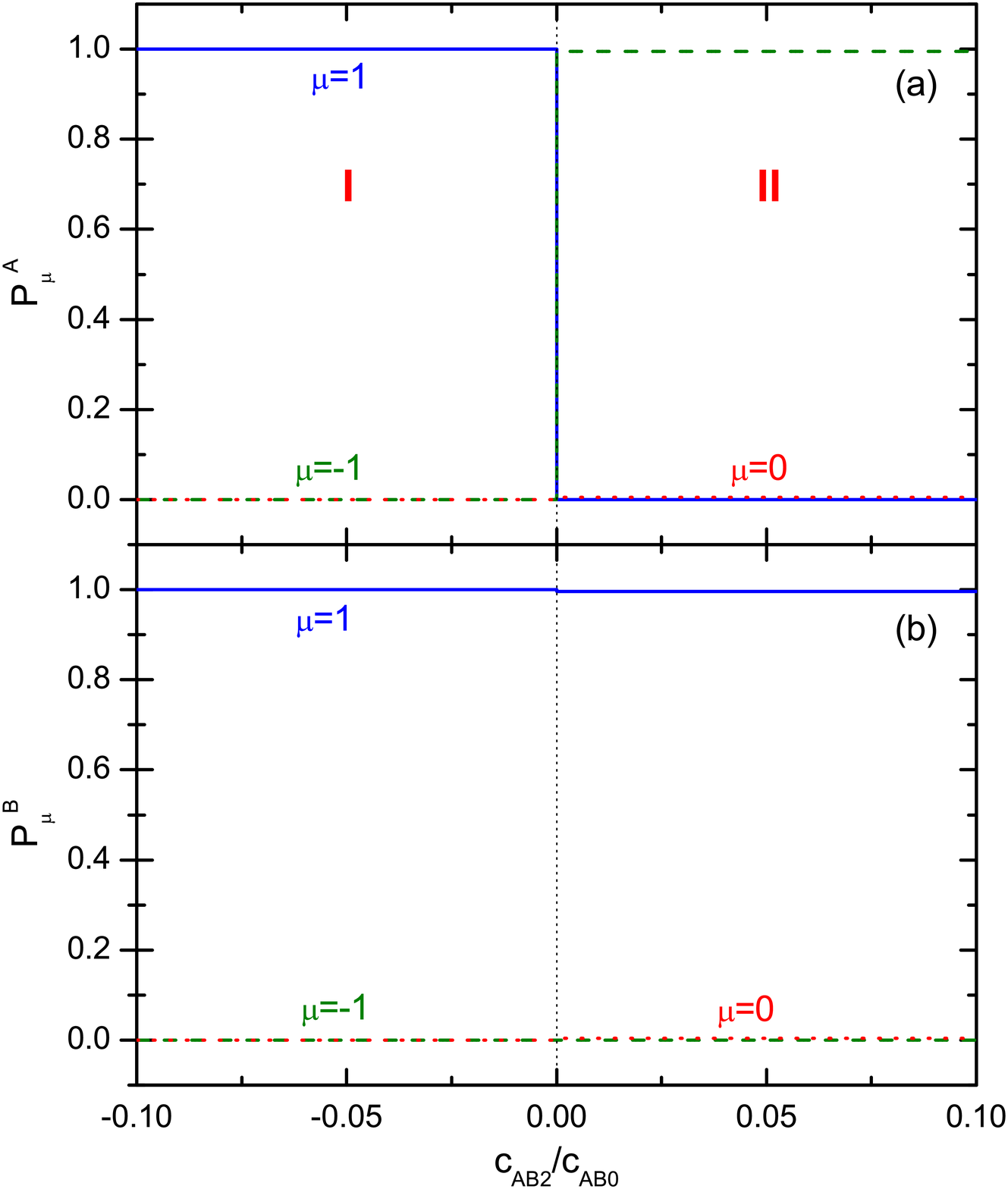}}
 \caption{(color online) $P_{\mu}^A$ (a) and $P_{\mu}^B$ (b) against $c_{AB2}/c_{AB0}$. The parameters are the same as in Fig.\ref{fig2}.}
 \label{fig6}
\end{figure}

\begin{figure}[htbp]
 \centering \resizebox{0.6\columnwidth}{!}{\includegraphics{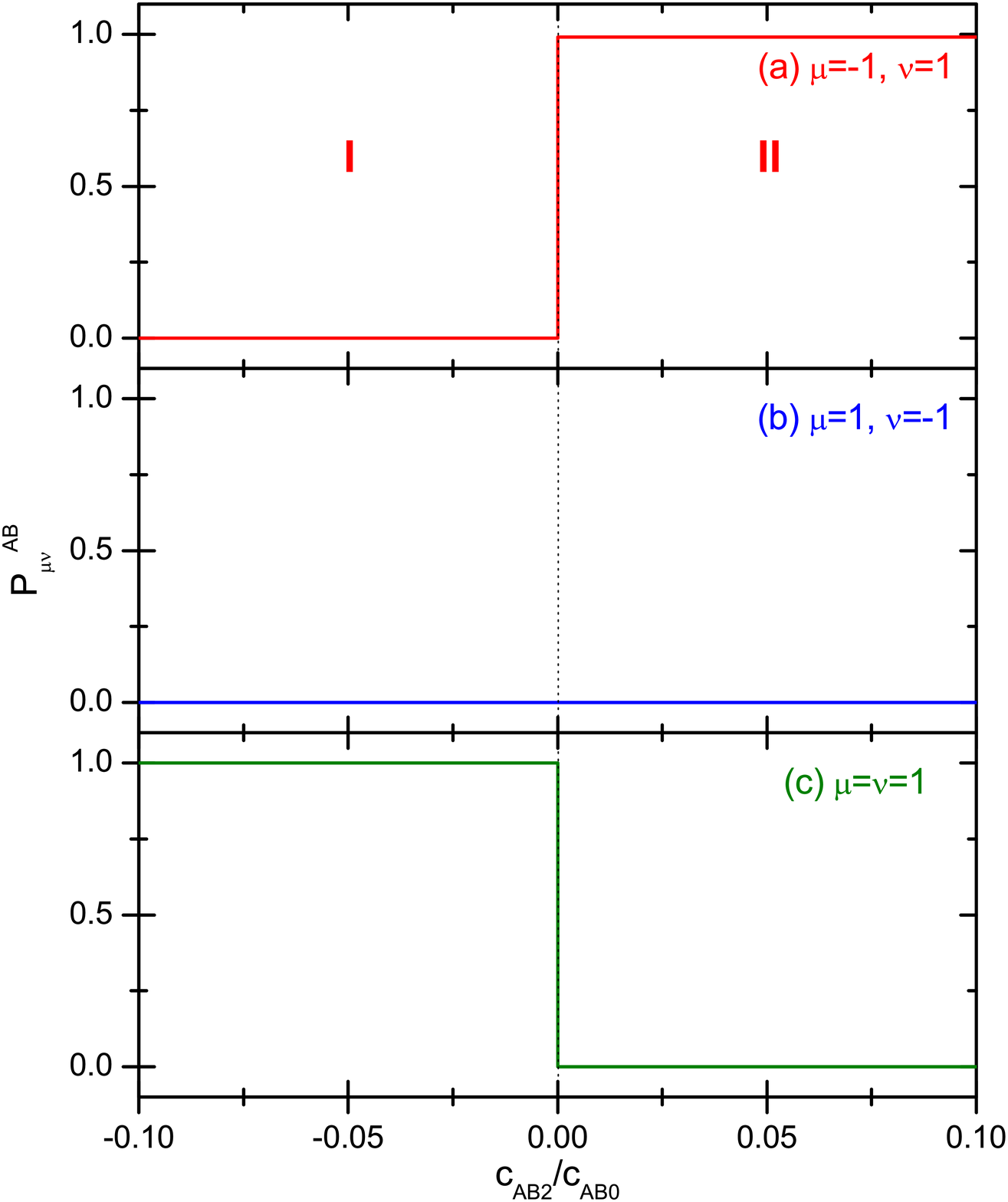}}
 \caption{(color online) Selected $P_{\mu\nu}^{AB}$ against $c_{AB2}/c_{AB0}$. The parameters are the same as in Fig.\ref{fig2}.}
 \label{fig7}
\end{figure}

\begin{figure}[htbp]
 \centering \resizebox{0.6\columnwidth}{!}{\includegraphics{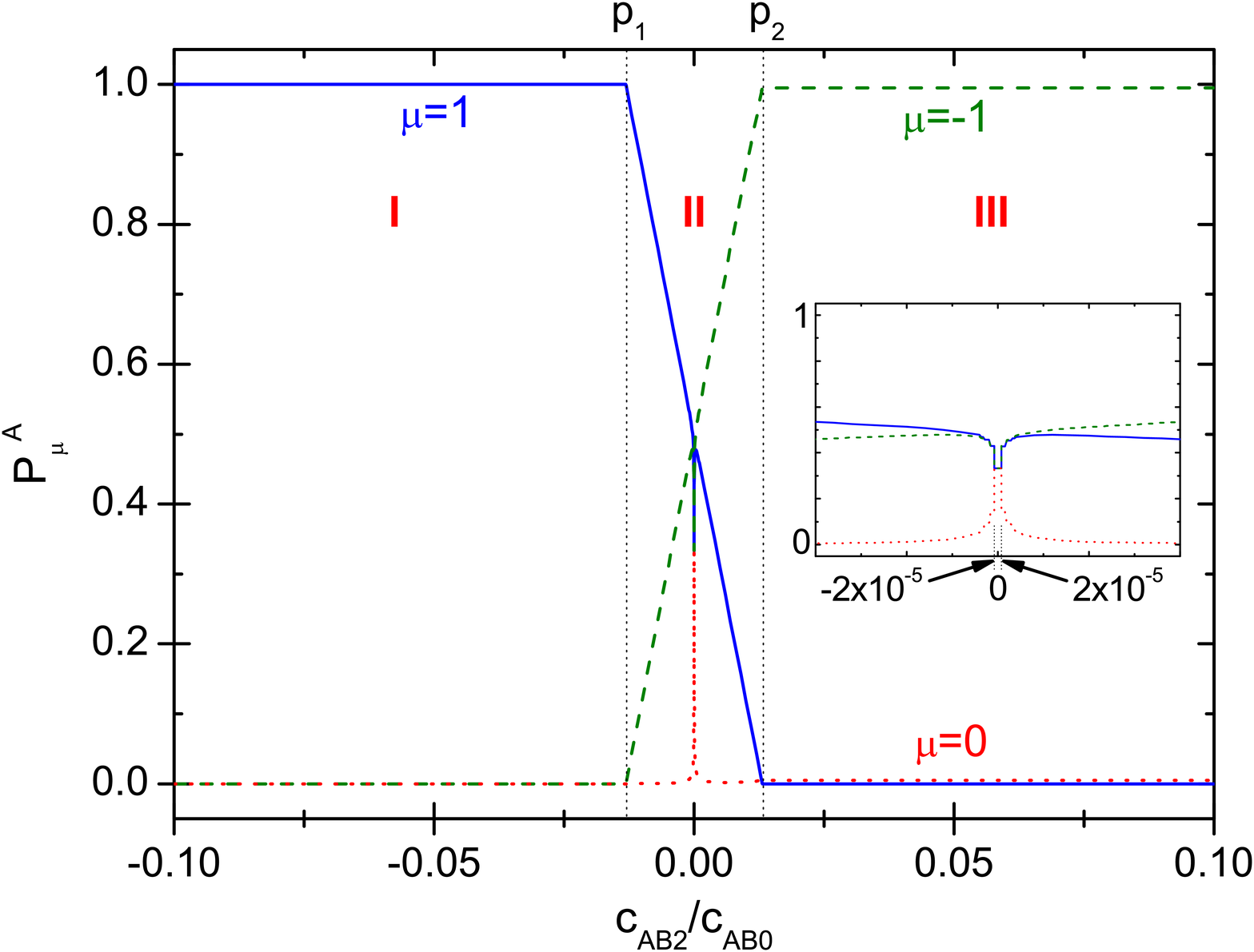}}
 \caption{(color online) $P_{\mu}^A$ against $c_{AB2}/c_{AB0}$. The parameters are the same as in Fig.\ref{fig3}. In this case $P_1^B=1$ disregarding $c_{AB2}$. }
 \label{fig8}
\end{figure}

\section*{Probability of an atom with spin-component $\mu$ and the probability of a pair of atoms with $\mu$ and $\nu$}

Since the spins are correlated in different ways in different spin-textures, the associated probability of an atom lying in a specific spin-component $\mu$ (denoted as $P_{\mu}^X$) and the probability of two atoms with their spins in $\mu$ and $\nu$ (denoted as $P_{\mu\nu}^{XX'}$) are also different in different spin-textures. Therefore, information on the spin-textures can be extracted from these probabilities. In theoretical derivation, a crucial point is to extract the spin-state of an atom or of two atoms from the total spin-state. By using the fractional parentage coefficients developed for spin-1 many-body systems \cite{bao05,bao06}, the extraction can be realized even when the details of the total spin-state have not yet been known. This is shown in the Appendix.

With these coefficients, the total spin-state $(\vartheta_{S_A}^{N_A}\vartheta_{S_B}^{N_B})_{SM}$ can be rewritten as
\begin{equation}
 (\vartheta_{S_A}^{N_A}\vartheta_{S_B}^{N_B})_{SM}
  =  \sum_{\mu}
     \chi_{\mu}(i)
     \sum_{S'}
     [ T_{\mu}^{S'}
       (\vartheta_{S_A+1}^{N_A-1}\vartheta_{S_B}^{N_B})_{S'M-\mu }
      +Q_{\mu}^{S'}
       (\vartheta_{S_A-1}^{N_A-1}\vartheta_{S_B}^{N_B})_{S'M-\mu} ],
 \label{tss}
\end{equation}
where $\chi_{\mu}(i)$, the spin-state of an $A$-atom in $\mu $ component, has been extracted.
\begin{eqnarray}
 &T_{\mu}^{S'}
  =  C_{1\mu;S',M-\mu}^{SM}
     a_{S_A}^{N_A}
     \sqrt{(2S_A+1)(2S'+1)}
     W(1,S_A+1,S,S_B;S_AS'),& \label{p} \\
 &Q_{\mu }^{S'}
  =  C_{1\mu;S',M-\mu }^{SM}
     b_{S_A}^{N_A}
     \sqrt{(2S_A+1)(2S'+1)}
     W(1,S_A-1,S,S_B;S_AS'),& \label{q}
\end{eqnarray}
where the Clebsch-Gordan coefficients and the W-coefficient of Racah have been introduced. Then, the normalization of the total wave function $\Psi_{S_AS_BSM}$ can be expanded as
\begin{equation}
 \langle
 \Psi_{S_AS_BSM}|
 \Psi_{S_AS_BSM}
 \rangle
  =  \sum_{\mu}
  P_{\mu}^A,
\end{equation}
where
\begin{equation}
 P_{\mu}^A
  =  \sum_{S'}
     [ (T_{\mu}^{S'})^2
      +(Q_{\mu}^{S'})^2],
\end{equation}
is just the probability of an $A$-atom with its spin lying at $\mu$. The probability of a $B$-atom at $\mu$, i.e. $P_{\mu}^B$, can be obtained simply by interchanging $N_A\leftrightarrow N_B$ and $S_A\leftrightarrow S_B$ in the above formulae.

When the coupled spin-state of two $A$-atoms has been extracted by using Eq.(\ref{tswf}), we have the expansion
\begin{equation}
 \langle
 \Psi_{S_AS_BSM}|
 \Psi_{S_AS_BSM}
 \rangle
  =  \sum_{\mu,\nu}
     P_{\mu\nu}^{AA},
\end{equation}
where
\begin{eqnarray}
 &P_{\mu\nu}^{AA}
  =  \sum_{S_A'S'}
     (X_{S_A'S'}^{\mu\nu})^2,& \\
 &X_{S_A'S'}^{\mu\nu}
  =  \sum_{\lambda}
     C_{1,\mu;1,\nu}^{\lambda,\mu+\nu}
     C_{\lambda,\mu+\nu;S',M-\mu-\nu}^{SM}
     h_{\lambda S_A'S_A}^{N_A}
     \sqrt{(2S_A+1)(2S'+1)}
     W(\lambda S_A'SS_B;S_AS'),&
\end{eqnarray}
where $\lambda$ runs over $0$ and $2$. $P_{\mu\nu}^{AA}$ is the probability of two $A$-atoms lying in spin-components $\mu$ and $\nu$, respectively. The probability $P_{\mu\nu}^{BB}$ can be similarly obtained by interchanging the indexes $A$ and $B$.

Let $i$ ($i'$) denotes an $A$-atom (a $B$-atom). When $\chi_{\mu}(i)$ and $\chi _{\nu }(i')$ have been simultaneously extracted, by using Eq.(\ref{fpc}) we have the expansion
\begin{equation}
 \langle
 \Psi_{S_AS_BSM}|
 \Psi_{S_AS_BSM}\rangle
  =  \sum_{\mu\nu}
     P_{\mu\nu}^{AB},
\end{equation}
where
\begin{eqnarray}
 &P_{\mu\nu}^{AB}
  =  \sum_{S_{AB}}
     [ (Z_{S_{AB}}^{I,\mu\nu})^2
      +(Z_{S_{AB}}^{II,\mu\nu})^2
      +(Z_{S_{AB}}^{III,\mu\nu})^2
      +(Z_{S_{AB}}^{IV,\mu\nu})^2],& \\
 &Z_{S_{AB}}^{I,\mu\nu}
  =  \sum_{\lambda}
     C_{1,\mu;1,\nu}^{\lambda,\mu+\nu}
     C_{\lambda,\mu+\nu;S_{AB},M-\mu-\nu}^{SM}
     a_{S_A}^{N_A}
     a_{S_B}^{N_B}
     \sqrt{(2\lambda+1)(2S_{AB}+1)(2S_A+1)(2S_B+1)}
     \left(
     \begin{array}{ccc}
      1      & 1      & \lambda \\
      S_A+1, & S_B+1, & S_{AB} \\
      S_A    & S_B    & S
     \end{array}
     \right),&
\end{eqnarray}
where $P_{\mu\nu}^{AB}$ is the probability of an $A$-atom and a $B$-atom lying in $\mu$ and $\nu$, respectively. The 9-$j$ symbol has been introduced, $\lambda$ runs over $0$, $1$, and $2$. $Z_{S_{AB}}^{II,\mu\nu}$ is similar to $Z_{S_{AB}}^{I,\mu\nu}$ but with $a_{S_B}^{N_B}$ being changed to $b_{S_B}^{N_B}$ and the index $S_B+1$ in the 9-$j$ symbol being changed to $S_B-1$. $Z_{S_{AB}}^{III,\mu\nu}$ is similar to $Z_{S_{AB}}^{I,\mu\nu}$ but with $a_{S_A}^{N_A}\rightarrow b_{S_A}^{N_A}$ and $S_A+1\rightarrow S_A-1$. $Z_{S_{AB}}^{IV,\mu\nu}$ is similar to $Z_{S_{AB}}^{I,\mu\nu}$ but with $a_{S_A}^{N_A}a_{S_B}^{N_B}\rightarrow b_{S_A}^{N_A}b_{S_B}^{N_B}$, $S_A+1\rightarrow S_A-1$, and $S_B+1\rightarrow S_B-1$.

Numerical examples of these probabilities are shown in Fig.\ref{fig4} to Fig.\ref{fig8}. Where $M=S$ is given, namely, the $Z$-axis is chosen lying along the orientation of $S$.

\section*{Variation of the spin-texture against the inter-species interaction}

Note that, when the spin-spin force is repulsive (attractive), two atoms with their spins anti-parallel (parallel) will be lower in energy. This point is important to the following analysis, where the unit for $c_{AB2}$ is $c_{AB0}$. For convenience, the spins of the $X$-atom is called $X$-spins.

(1) The case $c_{A2}>0$ and $c_{B2}>0$

Recall that, if $c_{AB2}=0$, both species are in polar phase, where the spins are two-by-two coupled to zero to form the singlet-pairs ($s$-pairs) in which the two spins are anti-parallel. Fig.\ref{fig1} demonstrates that, in the domain marked by zone III, $(S_A,S_B,S)=(0,0,0)$. Thus the polar-polar phase remains unchanged when $c_{AB2}$ is limited in zone III. On the other hand, Fig.\ref{fig4} demonstrates that all the $P_{\mu}^A=P_{\mu}^B=1/3$ in zone III. It implies isotropism. This is consistent with the texture $(0,0,0)$.

When $c_{AB2}\neq 0$, the optimized phase for $c_{AB2}$ alone is to have all the spins of the two species aligned along the same direction if $c_{AB2}<0$, or aligned along two opposite directions, one for a species, if $c_{AB2}>0$. Thus, when $|c_{AB2}|$ increases, the alignment of spins is inevitable.

(i) The case $c_{AB2}>0$

When $c_{AB2}$ enters into zone IV ($p_3<c_{AB2}<p_4$), from Fig.\ref{fig1} we found that $S_A=N_A$ (recall that $N_B/2<N_A<N_B$ is assumed, thus the spins of the species fewer in particle number are fully aligned). We also found $S_A>S_B$, and the magnitudes $S=S_A-S_B$. Since $S$ is given lying along the $Z$-axis, $S_A$ ($S_B$) must be lying along $+Z$-axis ($-Z$-axis) to assure $S=S_A-S_B$. Let the value of $S_B$ at $p_3$ be denoted by $S_B|_{p_3}$. The spin-texture-transition (STT) from $(0,0,0)$ to $(N_A,S_B|_{p_3},N_A-S_B|_{p_3})$ occurs at $p_3$. After the STT, the complete alignment of the $A$-spins is shown by Fig.\ref{fig4}a where $P_1^A=1$, $P_0^A=0$ and $P_{-1}^A=0$ in zone IV. The fact that $P_0^A=0$ demonstrates a complete breakdown of all the $s$-pairs of the $A$-spins. Meanwhile all the $B$-spins are also suddenly free from the $s$-pairs and re-align along the $\pm Z$-axis as shown in Fig.\ref{fig4}b where $P_0^B=0$ and $P_{-1}^B>1/2>P_1^B>0$ in the zone IV. Accordingly, $S_B=N_B(P_{-1}^B-P_1^B)$ lying along the $-Z$-axis. Remind that, for $c_{B2}$ ($>0$) alone, the phase with $P_{-1}^B=P_1^B=1/2$ would minimize the repulsion from $c_{B2}$; while for $c_{AB2}$ ($>0$) alone and when $P_1^A=1$, the optimized phase is to have $P_{-1}^B=1$. Thus the competing effects of $c_{B2}$ and $c_{AB2}$ lead to a balance so that $P_{-1}^B$ should be larger but not much larger than $1/2$. It results in $P_{-1}^B>1/2>P_1^B>0$ as found in Fig.\ref{fig4}b. In particular, when $N_A$ is sufficiently larger than $N_B/2$ (as we have assumed), $S_B$ will be smaller than $S_A$ when $c_{AB2}$ is not sufficiently large, but $S_B$ will exceed $S_A$ when $c_{AB2}$ increases further.

When $c_{AB2}$ increases from $p_3$, $P_{-1}^B$ increases linearly and $P_1^B$ decreases as shown in Fig.\ref{fig4}b. Accordingly, $S_B$ becomes larger and larger while $S=S_A-S_B$ becomes smaller. When $c_{AB2}=p_4$, the magnitudes $S_B=S_A$ and $S=0$, and the system as a whole suddenly becomes isotropic (Since $M=S$ is given, $S$ is lying along the $Z$-axis if $S\neq 0$. However, when $S=0$ the system is blind to the orientation. This leads to the isotropism). Accordingly, we found in Fig.\ref{fig4}a and \ref{fig4}b that all the six $P_{\mu}^X$ are suddenly changed to $1/3$ at $p_4$. It is noted that, during crossing over $p_4$, all the spins remain aligning but blind in direction. Say, the $A$-spins remain aligning along a direction so that $S_A=N_A$ remain unchanged as shown in Fig.\ref{fig1}, however they are no more lying along the $Z$-axis but arbitrary.

In zone V, $S_B>S_A$. Since $S_B$ and $S_A$ are lying along opposite directions, the overtaking of $S_B$ implies that $S$ has reversed its direction. Since $M=S$ is adopted, $S$ is always lying along the $Z$-axis. Thus the reverse of $S$ implies a reverse of the frame. Accordingly, although the alignment of the spins remain unchanged (say, $S_A=N_A$ holds), $P_{\mu}^A$ is changed from $\delta_{\mu,1}$ to $\delta_{\mu,-1}$ due to the reverse of frame. A similar change in $P_{\mu}^B$ occurs also. Then, $P_1^B$ keeps monotonically increasing and will arrive at 1 when $c_{AB2}=p_5$. Accordingly, $S_B=N_B(P_1^B-P_{-1}^B)$ keeps increasing and will be equal to $N_B$ when $c_{AB2}=p_5$. When $c_{AB2}$ crosses over $p_5$ and enters zone VI, all the $A$-spins ($B$-spins) are fully polarized along the $-Z$-axis ($+Z$-axis). Accordingly, $S=N_B-N_A$, and the associated texture is denoted as $(N_A,N_B,N_B-N_A)$ and is named anti-parallel f-f phase. In this phase, each $A$-spin is anti-parallel to every $B$-spins so that the repulsion from $c_{AB2}$ is minimized. Therefore, it is stable against the further increase of $c_{AB2}$. The turning point $p_5$ marks the boundary of this phase

(ii) The case $c_{AB2}<0$

Recall that the texture $(0,0,0)$ holds from $p_2$ to $p_3$. When $c_{AB2}$ enters into zone II, the breakdown of the $s$-pairs and the alignment of all the spins occur again. The $A$-spins align along a direction and the $B$-spins align along two directions as before. But the larger part of $B$-spins align now along the direction of the $A$-spins. Accordingly, $S_A=N_A$ and the magnitudes $S=S_A+S_B$. This is different from the texture in zone IV, where $S=S_A-S_B$. In zone II, $P_1^B$ increases with the decrease of $c_{AB2}$. When $c_{AB2}$ crosses over $p_1$ and enters into zone I, $P_1^B=1$ and each $B$-spin is lying parallel to every $A$-spins. Accordingly, $S_B=N_B$, and $S=N_A+N_B$ and the associated texture is denoted as $(N_A,N_B,N_A+N_B)$ and named parallel f-f phase. In this phase the attraction raised by $c_{AB2}$ is maximized so that it is stable when $c_{AB2}$ becomes more negative. The turning point $p_1$ marks the boundary of this phase.

Comparing Fig.\ref{fig1} and \ref{fig4}, we see that the STT are sensitively reflected in $P_{\mu}^X$. In particular, the five critical and/or turning points $p_1$ to $p_5$ are clearly shown in $P_{\mu}^X$. Besides, selected $P_{\mu\nu}^{AB}$ ($\mu$ is for an $A$-atom and $\nu$ is for a $B$-atom) are shown in Fig.\ref{fig5}. Information on the spin-textures can also be extracted from this figure. Say, all the $P_{\mu\nu}^{AB}=1/9$ in zone III and at the boundary separating Zone IV and V. This demonstrates the isotropism once $S=0$. Besides, $P_{-1,1}^{AB}=1$ in zone VI demonstrates directly the anti-parallel f-f phase. $P_{1,1}^{AB}=1$ in zone I demonstrates directly the parallel f-f phase. In zone II, $P_{1,1}^{AB}+P_{1,-1}^{AB}=1$ and $P_{1,1}^{AB}>P_{1,-1}^{AB}>0$ together demonstrate how the spins are aligned as stated above. Whereas, in zone IV, $P_{1,1}^{AB}+P_{1,-1}^{AB}=1$ and $P_{1,-1}^{AB}>P_{1,1}^{AB}>0$ together demonstrate the alignment of the spins as stated above.

(2) The case $c_{A2}<0$ and $c_{B2}<0$

It is shown in Fig.\ref{fig2} that, in this case, both species will keep the f phase disregarding how $c_{AB2}$ is. Nonetheless, the two groups of aligned spins will point to the same direction if $c_{AB2}<0$ (i.e., the g.s. has $S=N_A+N_B$ and is in parallel f-f phase), or point to opposite directions if $c_{AB2}>0$ (i.e., the g.s. has $S=|N_A-N_B|$ and is in anti-parallel f-f phase). This is confirmed by $P_{\mu }^X$ and $P_{\mu\nu}^{AB}$ as shown in Fig.\ref{fig6} and \ref{fig7}. Obviously, $c_{AB2}=0$ is a critical point where the transition $(N_A,N_B,N_A+N_B)\rightarrow (N_A,N_B,|N_A-N_B|)$ occurs.

(3) The case $c_{A2}>0$ and $c_{B2}<0$

It was found that $S_B=N_B$ (refer to Fig.\ref{fig3}) and $P_{\mu}^B=\delta_{\mu,1}$ hold in the whole range of $c_{AB2}$. Thus the $B$-spins are in the f phase lying along the $Z$-axis. This phase is very stable against $c_{AB2}$. Whereas, the $A$-spins are in the polar phase with $P_{\mu}^A=1/3$ when $c_{AB2}=0$ (refer to Fig.\ref{fig8}). However, the polar phase is extremely fragile against $c_{AB2}$. Once $|c_{AB2}|$ deviates from zero, $P_0^A$ falls from $1/3$ into zero in an extremely narrow domain surrounding $c_{AB2}=0$ (refer to the sub-figure inside Fig.\ref{fig8}). Remind that $P_0^A=0$ implies the breakdown of all the $s$-pairs. When $c_{AB2}$ increases further and arrives at $p_2$ (marked in Fig.\ref{fig8}), $P_{-1}^A$ increases from $1/3$ to 1 while $P_1^A$ decreases from $1/3$ to 0. Accordingly, $S_A$ increases from 0 to $N_A$ but lying along the $-Z$-axis opposite to the direction of $S_B$. This results in the magnitude $S=S_B-S_A$. When $c_{AB2}=p_2$, all the $A$-spins align along the $-Z$-axis. In this way each $A$-spin is antiparallel to every $B$-spin, and the mixture is in the antiparallel f-f phase. In this phase the repulsion from $c_{AB2}$ is minimized and therefore is stable against the further increase of $c_{AB2}$. Similarly, $p_1$ marks the boundary of the parallel f-f phase.

Comparing Fig.\ref{fig1} (where the zone III for the $(0,0,0)$ texture is broad) and Fig.\ref{fig3} (where $(0,0,0)$ exists only at a point), we know that the stability of the polar phase of the $A$-spins depends on the $B$-spins. If both are in polar phase, then both are stable until $|c_{AB2}|$ exceeds a critical value. Alternatively, if the $B$-spins are aligned, each $A$-spin will be attracted to align along the same direction (if $c_{AB2}<0$) or the opposite direction (if $c_{AB2}>0$) with them. This leads to the extremely fragility of the $(0,0,0)$ texture.

\section*{Summary}

We have solved the CGP numerically to obtain the solutions for 2-species spin-1 BEC, and we have calculated the probability densities $P_{\mu}^X$ and $P_{\mu\nu}^{AB}$ extracted from the exact solutions. The knowledge of the spin-textures is thereby obtained.

Remind that the g.s. of the 1-species spin-1 BEC has two well known phases, the polar and f phases. Thus, for 2-species, it is naturally to guess that there would be three phases: polar-polar, polar-f, and f-f phases. However, we know now:

(i) The polar-polar phase exists only if $c_{A2}>0$, $c_{B2}>0$, and $|c_{AB2}|$ is smaller than a critical value.

(ii) The presumed polar-f phase does not exist (unless $c_{AB2}=0$). This is because the polar phase will become extremely fragile when it is accompanied by an f phase even when $|c_{AB2}|$ is very weak. The aligned spins in f phase (say, the $B$-atoms) lure the spins in polar phase (say, the $A$-atoms) to align with them along the same or opposite direction (refer to Fig.\ref{fig3} and Fig.\ref{fig8}). Consequently, all the $s$-pairs of the $A$-atoms will be broken when $|c_{AB2}|$ deviates from zero. This is confirmed by the exact solution of the CGP, which leads to $P_0^A=0$. After being free from the $s$-pairs, the $A$-atoms will fall into a quasi-f (qf) phase in which the spins are divided into two parts lying parallel and anti-parallel to the direction of the $B$-spins (in f phase), respectively. When $c_{AB2}<0$ and becomes more negative, more and more $A$-spins are parallel to the $B$-spins. When all $A$-spins are parallel the mixture is in the parallel f-f phase. The case with $c_{AB2}>0$ is similar, but more $A$-spins are antiparallel to the $B$-spins. Thus the further increase of $c_{AB2}$ leads to the antiparallel f-f phase

(iii) The probability densities $P_{\mu}^X$ and $P_{\mu\nu}^{XX'}$ of each spin-texture have their own feature. Thus the theoretical calculation of $P_{\mu}^X$ and $P_{\mu\nu}^{XX'}$ as presented in this paper together with related experimental observation provides a way to discriminate the spin-textures.

(iv) The results from 2-species BEC give a hint for understanding multi-species BEC. For 3-species BEC, it is guessed that the polar-polar-polar could emerge only if all $c_{X2}$ ($X$ is for $A$, $B$ or $C$) are positive and the three $c_{XX'2}$ fall inside a limited scope, the polar-f-f phase does not exist but replaced by qf-f-f phase, and there would be f-f-f phase with two groups of spins lying along a direction while the third group lying reversely, and so on. These suggestions remain to be checked.

(v) Due to the existence of different phases, phase transitions are inevitable. It turns out that all the critical points where transition occurs are sensitive to the spin-dependent forces. These points can be theoretically predicted by calculating $P_{\mu}^X$ and $P_{\mu\nu}^{XX'}$, and can be experimentally identified via an observation on these probabilities. This provides a way for determining the strengths of these very weak forces.

\section*{Appendix}

Let $\vartheta_{S,M}^N$ be the normalized all-symmetric total spin-state of $N$ spin-1 atoms of the same kind, all the spins are coupled to $S$ and $M$. Then, $\chi(i)$ (the spin-state of the $i$-th particle) can be extracted as \cite{bao05,bao06}
\begin{equation}
 \vartheta_{SM}^N
  =  a_S^N
     [\chi(i)\vartheta_{S+1}^{N-1}]_{SM}
    +b_S^N
     [\chi(i)\vartheta_{S-1}^{N-1}]_{SM},
 \label{fpc}
\end{equation}
where $a_S^N$ and $b_S^N$ are the fractional parentage coefficients for extracting one particle. They appear as
\begin{eqnarray}
 &&a_S^N
  =  \sqrt{\frac{[1+(-1)^{N-S}](N-S)(S+1)}{2N(2S+1)}},  \nonumber \\
 &&b_S^N
  =  \sqrt{\frac{[1+(-1)^{N-S}]S(N+S+1)}{2N(2S+1)}},  \label{ab}
\end{eqnarray}
where $\vartheta_{S\pm 1}^{N-1}$ are also normalized and all-symmetric but for $N-1$ spins coupled to $S\pm 1$.

Let the $i$-th and $j$-th spins be coupled to $\lambda $, and the coupled spin-state is denoted by $[\chi (i)\chi (j)]_{\lambda }$. Then this pair can be extracted as
\begin{equation}
 \vartheta_{SM}^N
  =  \sum_{\lambda S'}
     h_{\lambda S'S}^N
     \{[\chi (i)\chi (j)]_{\lambda }\vartheta_{S'}^{N-2}\}_{SM},
 \label{tswf}
\end{equation}
where the coefficients $h_{\lambda S'S}^N$ are the fractional parentage coefficients for extracting two particles. They are
\begin{eqnarray}
 h_{0,SS}^N
 &=& [\frac{(N+S+1)(N-S)}{3N(N-1)}]^{1/2}, \nonumber \\
 h_{2,S+2,S}^N
 &=& [\frac{(S+1)(S+2)(N-S)(N-S-2)}{(2S+1)(2S+3)N(N-1)}]^{1/2}, \nonumber \\
 h_{2,S,S}^N
 &=& [\frac{S(2S+2)(N-S)(N+S+1)}{3(2S-1)(2S+3)N(N-1)}]^{1/2}, \nonumber \\
 h_{2,S-2,S}^N
 &=& [\frac{S(S-1)(N+S+1)(N+S-1)}{(2S-1)(2S+1)N(N-1)}]^{1/2}. \label{h2}
\end{eqnarray}
All the other $h_{\lambda S'S}^N$ are zero. Furthermore, when $S=0 $, $h_{2,S,S}^N$ and $h_{2,S-2,S}^N$ should be zero.

\section*{Acknowledgements}

Supported by the National Natural Science Foundation of China under Grants No.11372122, 11274393, 11574404, and 11275279; the Open Project Program of State Key Laboratory of Theoretical Physics, Institute of Theoretical Physics, Chinese Academy of Sciences, China(No.Y4KF201CJ1); the National Basic Research Program of China (2013CB933601); and the Natural Science Foundation of Guangdong of China (2016A030313313).

\section*{Author contributions}

Y. Z. He is responsible to the numerical calculation. Y. M. Liu is responsible to the theoretical derivation. C. G. Bao provides the idea, write the paper, and responsible to the whole paper. All authors reviewed the manuscript.

\section*{Additional information}

\textbf{Competing Interests:} The authors declare that they have no competing interests.

\end{document}